\documentclass[pdflatex,sn-mathphys-num,iicol]{sn-jnl}

\usepackage{graphicx}%
\usepackage{multirow}%
\usepackage{amsmath,amssymb,amsfonts}%
\usepackage{amsthm}%
\usepackage{mathrsfs}%
\usepackage[title]{appendix}%
\usepackage{textcomp}%
\usepackage{manyfoot}%
\usepackage{booktabs}%
\usepackage{algorithm}%
\usepackage{subcaption}
\usepackage{algorithmicx}%
\usepackage{algpseudocode}%
\usepackage{listings}%
\usepackage{cleveref}

\usepackage{makecell}
\usepackage[table,xcdraw]{xcolor}

\begin{document}

\title{Diffusion Transformers with Hybrid Conditioning for Structural Optimization }


\author[1]{\fnm{Aaron} \sur{Lutheran}}

\author[2]{\fnm{Srijan} \sur{Das}}

\author*[1,3]{\fnm{Alireza} \sur{Tabarraei}}\email{atabarra@charlotte.edu}

\affil[1]{\orgdiv{Department of Mechanical Engineering and Engineering Science}, \orgname{The University of North Carolina at Charlotte}, \orgaddress{\city{Charlotte}, \state{NC}, \postcode{28223}, \country{USA}}}

\affil[2]{\orgdiv{Department of Computer Science}, \orgname{The University of North Carolina at Charlotte}, \orgaddress{\city{Charlotte}, \state{NC}, \postcode{28223}, \country{USA}}}

\affil[3]{\orgdiv{School of Data Science}, \orgname{The University of North Carolina at Charlotte}, \orgaddress{\city{Charlotte}, \state{NC}, \postcode{28223}, \country{USA}}}


\abstract{This work presents a diffusion transformer framework for data‑driven structural topology optimization that combines the accuracy of physics‑based methods with the efficiency of generative deep learning. Conventional approaches such as the Solid Isotropic Material with Penalization (SIMP) method require repeated finite element analyses at every iteration, making large‑scale or real‑time optimization computationally expensive. We propose a hybrid conditioning diffusion transformer (DiT) model that learns to generate near‑optimal topologies directly from problem definitions, eliminating iterative analysis during inference. The model integrates spatially distributed conditioning through concatenated stress and strain fields and global conditioning via adaptive layer normalization (AdaLN) using scalar descriptors such as load position, magnitude, and prescribed volume fraction. A dataset of 30,000 two‑dimensional SIMP‑optimized structures was generated for training and evaluation. Results demonstrate that the proposed DiT achieves less than 1\% compliance errors relative to ground‑truth SIMP solutions while maintaining accurate volume fractions and structural connectivity. Deterministic DDIM sampling enables high‑fidelity topology generation in seconds using as few as five denoising steps, enabling near-real‑time performance. The hybrid conditioning diffusion transformer thus provides an efficient and scalable alternative to traditional topology optimization methods, with strong potential for integration into interactive computer‑aided design workflows.}


\maketitle

\section{Introduction}\label{sec: Intro}

Engineering design requires the consideration of many possible solutions in design spaces which have many independent variables. With design at scale, it becomes important to develop methods which can discover efficient solutions automatically, rather than relying on intuition or trial and error. With the improvement of computational resources, optimization techniques have gained significant traction and development. These methods narrow down large solution spaces to only a handful of options that engineers can evaluate, choose, and modify. Topology optimization (TO) is one such technique for creating efficient structures in mechanical, aerospace, and civil engineering applications \cite{rong_structural_2022}. It offers a framework for determining material layouts in a design domain to achieve the best performance.

For structural TO, the Solid Isotropic Material with Penalization (SIMP) method is one of the most widely adopted techniques due to its robustness and adaptability \cite{sigmund_topology_2013}. In the SIMP method, the design domain is divided into finite elements, where each element is treated as its own design variable. Each design variable controls the material properties of its element, ranging from a fully solid to fully void material. Each design variable is independent, which leads to a vast design space that introduces little bias on the final optimized design \cite{bendsoe_topology_2004, andreassen_efficient_2011, bendsoe_optimal_1989}. Minimum finding methods can then be implemented on these design variables, converging on solutions which optimize for design metrics, such as structural compliance, thermal transfer, or electro-magnetic behavior.

Despite the success of SIMP-based topology optimization schemes, the method faces several limitations. SIMP requires a finite element analysis (FEA) at every iteration to evaluate responses and sensitivities, which can be computationally expensive for large-scale or high-resolution problems. Expanding the problem into 3-dimensional domains compounds the problem, requiring significant compute resources \cite{behzadi_real-time_2021}. Convergence can be sensitive to mesh resolution, boundary conditions, and load configurations \cite{zhu_filter_2015, wang_projection_2011, white_computational_2019}, requiring multiple runs to determine valid design options. These factors make real-time or interactive design applications impractical using conventional optimization alone \cite{behzadi_real-time_2021, lazarov_length_2016}.

To address some of these challenges, alternative formulations have been proposed over the years. Level-set methods represent material boundaries implicitly through continuous scalar fields and evolve them using shape derivatives. These methods naturally produce smooth boundaries but often require reinitialization steps and can struggle with topological changes such as merging or splitting of features \cite{allaire_level-set_2002, luo_level_2008}.

Gradient-free approaches, such as genetic algorithms, particle swarm optimization, or simulated annealing, have also been explored for topology design problems where gradient information is unavailable or unreliable \cite{hajela1993genetic, luh2011binary, cui2002topology, garcia2011hybrid}. However, their high computational cost typically restricts their use to small-scale problems or coarse discretizations \cite{tang_topology_2024, sigmund_topology_2013}. Compared with these alternatives, SIMP remains attractive because it provides a differentiable framework compatible with efficient gradient-based solvers, even though it remains computationally demanding.

Recent advances in machine learning (ML) have inspired new approaches that aim to bypass iterative FEA-based optimization by learning direct mappings from problem definitions to optimal topologies. Convolutional neural networks (CNNs), U-Net architectures, generative adversarial networks (GANs), and variational autoencoders (VAEs) have all been investigated for predicting near-optimal designs from boundary conditions and load distributions \cite{chandrasekhar_tounn_2021, nie_topologygan_2021, zheng_generating_2021, banga_3d_2018, shin_topology_2023, shishir_multimaterials_2024, behzadi_real-time_2021, SHISHIR2024107218}. Of these approaches, the most success has been found in approaches which implement diffusion or latent diffusion architectures \cite{giannone_diffusing_2023, maze_diffusion_2022, zhang_research_2025, lutheran_latent_2025}.

A key insight in this line of research is that incorporating physics-informed features significantly improves predictive accuracy. Stress and strain fields derived from initial FEA encodes essential information about load paths and deformation patterns within the structure. These fields act as priors that guide ML models toward physically plausible solutions without requiring explicit FEA during inference \cite{nie_topologygan_2021, jeong_physics-informed_2023}.

Performance evaluation in ML-based topology prediction typically involves metrics such as compliance error percentage relative to ground-truth SIMP results, deviation in prescribed volume fraction, and connectivity. While many models achieve visually convincing results with low compliance errors on test cases similar to training data, they often struggle with generalization across unseen boundary conditions or load directions \cite{maze_diffusion_2022}. Furthermore, generated structures may contain disconnected regions or unrealistic geometries that would be infeasible in practice. Additionally, integrating trained models into computer-aided design (CAD) environments requires low-latency inference and highly trustworthy designs \cite{behzadi_real-time_2021, lutheran2026physics}. Balancing a model's representational capacity with runtime efficiency remains an open challenge in practical ML-driven topology optimization workflows.

Diffusion-based generative models have recently demonstrated success in producing high-fidelity images across multiple domains. The denoising diffusion probabilistic model (DDPM) framework formulates generation as a two-step stochastic process: a forward diffusion process gradually corrupts data by adding Gaussian noise over many timesteps, while a reverse process learns to iteratively denoise samples back toward clean data distributions. The reverse process is parameterized by a machine learning model trained to predict the noise added at each step. Once trained, sampling proceeds by iteratively applying the learned denoising model starting from pure Gaussian noise until all noising timesteps are reversed. The representative capacity of these diffusion based models is higher than other machine learning image generation techniques and leads to better image quality \cite{nichol_improved_2021, yang_diffusion_2023, song_solving_2023, dhariwal_diffusion_2021, rombach_high-resolution_2022}. 


Deterministic variants such as Denoising Diffusion Implicit Models (DDIMs) modify this reverse process by removing the probabalistic component of the noising process while preserving consistency with DDPM trajectories \cite{song_denoising_2022, salimans_progressive_2022, yang_diffusion_2023}. This allows generation using fewer timesteps, often reducing the number of sampling steps from thousands down to tens of steps. This significantly reduces inference time without major degradations in sample quality.

While early diffusion models employed convolutional U-Nets as denoising backbones due to their strong local feature extraction capabilities, recent work has shown that replacing CNNs with Transformer architectures yields substantial gains in flexibility and scalability. Diffusion Transformers (DiTs) treat images as sequences of patch embeddings processed through self-attention layers capable of modeling long-range spatial dependencies across an entire domain \cite{peebles_scalable_2023}. These models are more data efficient and have better scaling properties than their U-Net based counterparts.

Additionally, transformers have a larger receptive field than CNN models, as every patch of the input domain is able to interact with every other patch. The capacity for transformers to learn long range relationships has promise for machine learning based topology optimization, as distant constraints in the design problem can influence local material placement. Furthermore, transformer architectures accomodate multiple conditioning mechanisms through token-level conditioning or adaptive normalization methods, making them well suited for the multi-modal constraints structural design problems pose \cite{croitoru_diffusion_2023, hong_improving_2023, dhariwal_diffusion_2021, guo_adaln_2022}.

In this work, we adopt a hybrid conditioning strategy combining both spatially distributed conditioning via concatenation of stress-strain fields and global modulation through adaptive layer normalization (AdaLN) based on scalar problem descriptors \cite{guo_adaln_2022}. This combination allows our diffusion transformer framework to capture both local mechanical behavior encoded in field data and global design constraints influencing overall topology distribution.

\section{Topology Optimization}\label{sec: Topopt}
\begin{figure}
    \centering
    \includegraphics[width=1\linewidth, trim={10mm, 6mm, 10mm, 6mm}, clip]{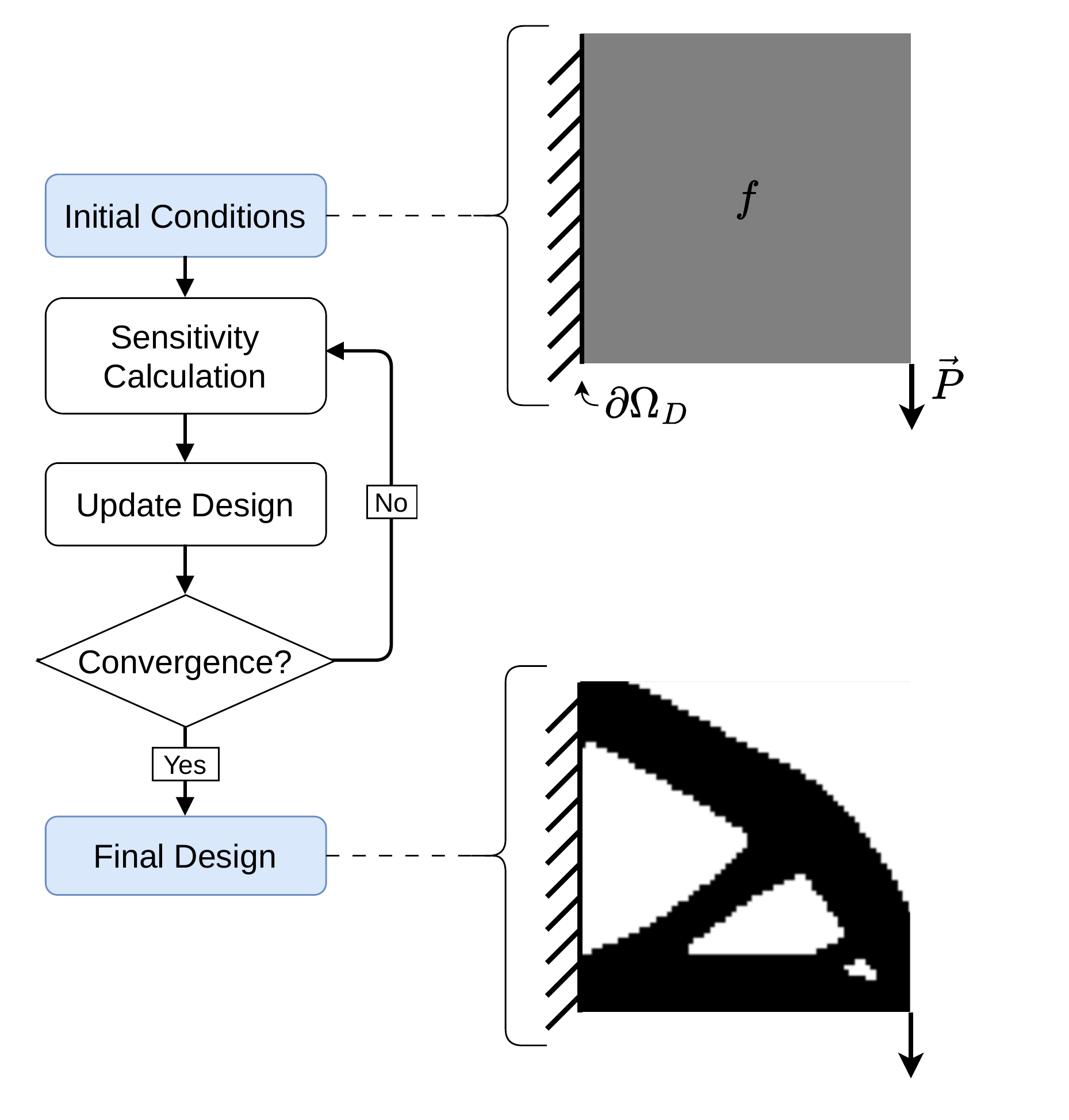}
    \caption{Example of a structural topology optimization problem. A prescribed volume fraction $f$ is present for the whole domain. A Dirichlet boundary condition exists along the domain boundary $\partial\Omega_D$ along with a load $\vec{P}$. The topopt algorithm returns a local optimized topology.}
    \label{fig:topopt_ex}
\end{figure}

There are many techniques for determining the best possible distribution of material for a set of design constraints. To evaluate which design performs the best, an objective function is needed to quantify the structure’s performance. If this objective function is differentiable, then a first order update scheme can be used to find a local minimum of the objective function, determining an optimal topology.

For structural problems involving force loads and boundary conditions, the most common objective function is the minimum compliance objective, which measures the total strain energy in the system. Other types of problems can also be solved using topology optimization by choosing different objectives, such as minimizing weight, maximizing natural frequency, or optimizing thermal or fluid flow properties. \Cref{fig:topopt_ex} depicts an example of a structural optimization problem and the conditions that govern the problem definition.

The SIMP method is a framework for density-based topology optimization. In this approach, the design variable represents a continuous material density field, where each point in the domain (or each finite element) takes on a density value
\begin{equation}
\rho \in [0, 1],
\label{eqn:rho_range_TO}
\end{equation}
where $\rho = 1$ represents solid material and $\rho = 0$ represents void.

Intermediate values of $\rho$ are allowed to make the problem continuous but are penalized during optimization to encourage discrete solid-void solutions. Standard minimization methods can then be used to find the set of densities $\rho$ that minimize the objective function.

In the SIMP approach, each finite element’s stiffness is scaled according to its material density through a power-law relationship
\begin{equation}
E_e = E_{min} + \rho_e^p(E_0-E_{min}),
\label{eqn:power_law_TO}
\end{equation}
where $E_e$ is the effective Young’s modulus of element $e$, $E_0$ is the Young’s modulus of the solid material, $E_{min}$ is the minimum Young's modulus used to prevent inversion errors, $p$ is the penalization factor (typically chosen as 3), and $\rho_e$ is the density of element $e$.

This interpolation ensures that elements with intermediate densities contribute less efficiently to stiffness compared to fully dense elements. This technique discourages gray regions and promotes discrete binary solutions for the density at every finite element in the structure. To implement structures generated by topology optimization, a thresholding step is applied to eliminate any remaining intermediate densities in the final topology.

The compliance minimization is evaluated using finite element analysis by computing the total strain energy stored in the structure, equivalent to the work done by external forces during deformation. The mathematical formulation of the compliance minimization problem can be written as

\begin{align}
\min_{\rho} \quad & C(\rho) = \mathbf{U}^T \mathbf{K}(\rho) \mathbf{U} \\
\text{subject to:} \quad 
& \mathbf{KU}=\mathbf{F}, \\
& V(\rho) / V_0 = f, \\
& 0 \leq \rho_e \leq 1,
\label{eqn:compliance_problem_TO}
\end{align}

where $C(\rho)$ is the total structural compliance, $\mathbf{K}(\rho)$ is the global stiffness matrix $\mathbf{U}$ is the global displacement vector obtained from equilibrium equations, $f$ is the prescribed volume fraction, and $V_e$, $V_0$ are the element and total domain volumes respectively.

The constraints ensure that the prescribed volume fraction is maintained and each design variable remains within its allowable bounds.

To determine the topology that best satisfies the compliance minimization objective, a gradient descent update scheme is used to iteratively adjust the material distribution. The derivative of the compliance function with respect to the element densities quantifies how changes in each element’s density affect overall structural performance. These sensitivities are then used to update the topology until a local minimum is reached. The derivative of compliance $C$ with respect to an element density $\rho_e$ can be derived using the chain rule

\begin{equation}
\frac{\partial C}{\partial \rho_e} = -\mathbf{U}^T \frac{\partial \mathbf{K}}{\partial \rho_e} \mathbf{U}
\end{equation}

Using the element stiffness equation to determine the gradient of the stiffness matrix, then substituting into the compliance derivative yields

\begin{equation}
\frac{\partial C}{\partial \rho_e} = -p \rho_e^{p-1} \mathbf{U}_e^T \mathbf{K}_e \mathbf{U}_e
\end{equation}

This sensitivity value for each element reflects how important that material region is in reducing deformation. Large sensitivities indicate that elements are under high stress or strain energy; increasing their density improves stiffness. Small sensitivities indicate that elements carry little load, so reducing their density has minimal impact on global stiffness.

The update rule for topology optimization using SIMP can be expressed as:

\begin{equation}
\rho_e^{(k+1)} = 
  \max(\rho_{\min}, 
  \max(0, 
  \min(1, 
  B_e)))
\end{equation}
where

\begin{equation} B_e = 
\rho_e^{(k)} 
  m
  (-\frac{\partial C}{\partial\rho_e}/\lambda)^\eta
  , 
  m,\eta>0
  , 
  \lambda
  \end{equation}
  
The Lagrange multiplier ($\lambda$) is adjusted at each iteration to satisfy the volume constraint:

\begin{equation}
V^* = \sum_{e=1}^{N} v_e\, \rho_e - V_{\text{allowed}} = 0
\end{equation}

Convergence can be satisfied when either the relative change in compliance between iterations falls below a specified tolerance or a maximum number of iterations has been reached. The resulting topology after convergence represents an optimal distribution of material that minimizes compliance while satisfying volume constraints.

\section{Diffusion Transformers}\label{sec: DiT}

DiTs are a class of generative models that combine the representational capacity of diffusion-based methods with the scalability and flexibility of Transformer architectures. The diffusion framework decomposes image generation into two processes: a forward diffusion process, which gradually corrupts data with noise, and a reverse denoising process, which learns to invert that corruption.

In the forward process, Gaussian noise is incrementally added to an image over a sequence of $T$ timesteps. Each step removes a small amount of information from the data, producing progressively noisier versions of the original image. Formally, the forward transition is defined as

\begin{equation}
q(x_t|x_{t-1}) = \mathcal{N}\!\left(x_t;\sqrt{1-\beta_t}\,x_{t-1},\,\beta_t I\right),
\end{equation}

where $\beta_t$ is a variance schedule controlling how much noise is introduced at each step. After $T$ steps, the marginal distribution can be expressed as

\begin{align}
q(x_t|x_0) &= \mathcal{N}\!\left(x_t;\sqrt{\bar{\alpha}_t}x_0,(1-\bar{\alpha}_t)I\right), \\
\alpha_{t} &= 1-\beta_{t}, \\ \bar{\alpha}_{t}&=\prod_{s=1}^t\alpha_{s}.
\end{align}

This formulation allows direct sampling of any intermediate noisy state $x_t$ without iteratively applying each previous step, which simplifies training and analysis.

The goal of training is to approximate the reverse conditional distribution that reconstructs cleaner samples from noisier ones

\begin{equation}
p_\theta(x_{t-1}|x_t,c) =
\mathcal{N}\!\left(x_{t-1};\, 
    \mu_\theta(x_t,t,c),\, 
    \Sigma_\theta(x_t,t,c)
\right),
\end{equation}

where $c$ represents optional conditioning variables. The model learns to predict either the mean or the noise component added at each step. This approach reframes image generation as a sequence of small denoising tasks rather than a single complex synthesis operation, which stabilizes training and improves sample diversity.

Conditional diffusion models extend this process by incorporating external information, such as class labels, text embeddings, or other modalities, to guide denoising. The conditioning signal remains uncorrupted throughout diffusion and is available at every timestep.

In DiTs, the denoising network is implemented using a Transformer rather than a convolutional U-Net. The input image is divided into non-overlapping patches of size $p \times p$, each flattened and linearly projected into an embedding vector of dimension $d$. These patch embeddings serve as tokens for the Transformer.

Within each Transformer block, multi-head self-attention computes dependencies between tokens
\begin{align}
Q &= XW_Q,\\ K&=XW_K,\\ V&=XW_V,\\
\text{Attention}(Q,K,V)
&=\text{softmax}\!\left(\frac{QK^T}{\sqrt{d_k}}\right)V.
\end{align}

where $W_Q, W_K,$ and $W_V$ are learned weight matrices for the queries, keys, and values. This mechanism allows the model to capture long-range spatial relationships across the image domain---something less efficiently handled by convolutional architectures. A pointwise feedforward network follows attention layers to introduce nonlinearity and further mix information across channels.

\subsection{Conditioning Mechanisms}

Conditioning information $c$ can be integrated into DiTs through several strategies. In-context conditioning is a technique where the conditioning signal is projected into an additional token with the same embedding dimension as image tokens. This token participates in self-attention alongside image tokens, allowing contextual integration without modifying the attention structure.

Cross-Attention introduces a separate attention mechanism that uses queries derived from image tokens, with keys and values derived from conditioning tokens. This integrates features between conditioning sources and is used to introduce conditioning that comes a different data modality. For example, text conditioning image generation. As long as the text can be mapped to a token sequence, it can be projected into the keys and values for the cross-attention mechanism, informing the image generation.

Adaptive Layer Normalization (adaLN) introduces scale and shift parameters in layer normalization are modulated by functions of the conditioning information $c$. These parameters are typically produced by a small MLP with $c$ as the input, allowing fine-grained control over feature transformations within both the attention and feedforward blocks.

Each method provides different trade-offs between computational cost and conditioning strength. AdaLN is lightweight and effective for continuous conditioning signals, while cross-attention better handles structured or multimodal inputs. In-context conditioning allows the model to inform tokens without adding the overhead of additional attention blocks.

\subsection{Inference for Diffusion Models}

During inference, deterministic sampling methods such as DDIM can accelerate generation while maintaining high fidelity. Instead of performing all $T=1000$ diffusion steps used during training, DDIM allows sampling at a subset of timesteps while maintaining consistent noise scheduling under linear variance assumptions. The DDIM update rule modifies the reverse process to reuse predicted noise terms across steps, reducing stochasticity and computational load.

In the standard diffusion framework, the reverse process determines the mean $\mu_{\theta}$ and deviation $\sigma_t$ to update the sample

\begin{align}
\mu_\theta(x_t,t) &= 
    \frac{1}{\sqrt{\alpha_t}}
    \left(
        x_t - 
        \frac{1 - \alpha_t}{\sqrt{1 - \bar{\alpha}_t}}
        \epsilon_\theta(x_t, t, c)
    \right)\\
\sigma_t^2 &= \frac{1-\bar{\alpha}_{t-1}}{1-\bar{\alpha}_{t}}\beta_t\\
x_{t-1} &= 
    \mu_\theta(x_t,t)
    + \sigma_t z
\end{align}

Where $z\sim\mathcal{N}(0,\mathbf{I})$ is Gaussian noise and $\epsilon_\theta$ is the noise as predicted by the model. The introduction of Gaussian noise in denoising adds a stochastic component that introduces an uncertainty at every timestep. DDIM removes this stochastic term, which yields a deterministic mapping between successive timesteps

\begin{equation}
x_{t-1} =
    \sqrt{\bar{\alpha}_{t-1}}\,x_0
    + 
    \sqrt{1 - \bar{\alpha}_{t-1}}\,
    \epsilon_\theta(x_t, t, c),
\end{equation}

DDIM allows intermediate steps to be deterministically interpolated along this trajectory. This allows for subsampling, where instead of performing all $T=1000$ denoising steps, inference can be done over a reduced set of timesteps (e.g., using 50--100 instead of 1000). High-quality samples can often be produced with 10--20 times fewer forward passes through the denoising model. Excessive subsampling can degrade sample fidelity due to insufficient refinement in later steps. 

\section{Methodology}\label{sec: Methodology}
\subsection{Dataset Generation}\label{subsec: Dataset}
\begin{figure*}
    \centering
    \includegraphics[width=0.9\linewidth]{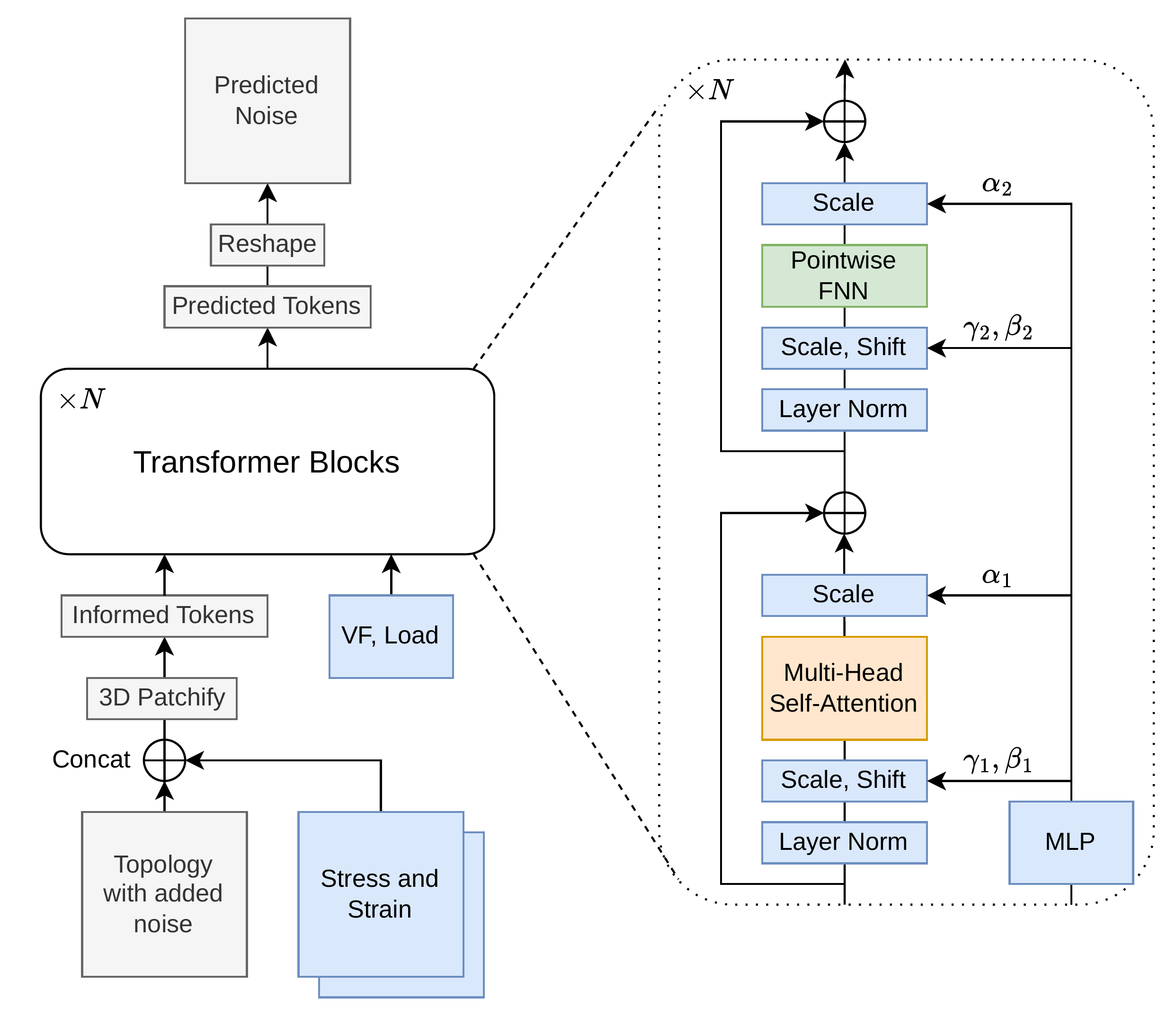}
    \caption{Diagram of conditioning used for the topology optimization DiT model, using a hybrid conditioning scheme. Global variables are passed into the model with adaLN conditioning, while the stress and strain input fields are concatenated to the topology input channel.}
    \label{fig:hybrid}
\end{figure*}

In order to implement the proposed diffusion transformer framework for two-dimensional structural topology optimization, a dataset was generated using finite topology optimization. The dataset consists of a set of optimized topologies on a two-dimensional square domain discretized into $64 \times 64$ square elements. Boundary conditions were randomly assigned for each sample to ensure diversity in problem definitions. Between 1 and 4 boundary constraints were applied as either fixed points in the corners or midpoints of edges, or as fixed segments connecting these locations. A single external load was applied along the edge of the domain per sample. The load applied has unit magnitude with a direction sampled randomly. The target material volume fraction $f$ was randomly sampled from a uniform distribution in the range $0.3$ to $0.5$.

A total of 30,000 optimized samples were generated using SIMP topology optimization with the process described in \cref{sec: Topopt}. A penalization factor of 3 was used with 100 optimization steps as the stopping criterion. The dataset was divided into two splits, reserving 90\% for training and 10\% for validation. To contribute to the conditioning, the von Mises stress and strain energy density were calculated on the initial domain assuming full material presence. This encodes the boundary conditions and their relationship to the load in a dense image field rather than a sparse coordinate space along the boundary of the domain.

Each sample consists of:
\begin{itemize}
    \item   A binary topology map,
    \item   The von Mises stress field,
    \item   The strain energy density field,
    \item   The load coordinates and magnitude components, and
    \item   The prescribed volume fraction.
\end{itemize}

No normalization was applied to the stress or strain energy values to preserve their magnitudes in relation to other samples.

\subsection{DiT Model Configuration}\label{sec: Model Config}
\begin{table}[]
\resizebox{0.95\linewidth}{!}{%
\begin{tabular}{c|ccc}
\multicolumn{1}{l}{\textbf{Model}} &

  \multicolumn{1}{l}{\textbf{Depth}} &
  \multicolumn{1}{l}{\textbf{Token Dim}} &
  \multicolumn{1}{l}{\textbf{Heads}} \\ 
  \hline
DiT-B & 12 & 768 & 12 \\
DiT-S & 12 & 384 & 6  \\
DiT-T & 8  & 192 & 3 
\end{tabular}%
}
\caption{List of model parameters for the DiT models. Models of size base (B), small (S), and tiny (T) are used. The final number in a model name indicates patch size. e.g. DiT-S-2 is a DiT small with a patch size of 2.}
\label{tbl:model_sizes}
\end{table}

The goal of the DiT is to learn the relationship between conditioning variables (comprising load information, volume fraction, and stress and strain fields) and the optimal topology distribution.

Because a uniform grid of square elements is used for the topology optimization scheme, each optimal topology can be stored as a $64 \times 64$ resolution image. The same applies to both von Mises stress and strain energy density fields.

Two types of conditioning are used in the DiT model: in-context conditioning and AdaLN conditioning. Traditional in-context conditioning is performed by concatenating an additional conditioning token to the set of tokens that represent the noisy data. Rather than encoding all conditions into a single token, we perform a channel-wise concatenation of the stress and strain conditioning fields with the noisy topology field. This allows positional relationships between stress/strain and topology to be preserved by embedding this information into each token based on its position. \Cref{fig:hybrid} depicts the conditioning method used for the model.

Each input sample is represented as a three-channel tensor. Channel 1 contains the topology, while channels 2 and 3 are the conditioning stress and strain fields. Noise is applied only to the topology field since both von Mises stress and strain energy density are known fields and are deterministic from the problem definition.

AdaLN conditioning is used to incorporate global information that is not spatially distributed across pixels. A conditioning vector
\begin{equation}
c = [x_{\text{load}}, y_{\text{load}}, F_x, F_y, v_f]
\end{equation}
is created from this global information. This vector is passed through a multilayer perceptron (MLP) to determine scale and shift factors $\alpha_1$, $\beta_1$, $\gamma_1$, $\alpha_2$, $\beta_2$, $\gamma_2$. These parameters globally influence all tokens within a given representation, allowing factors such as volume fraction to modulate material presence across all pixels.

The model is trained using three different patch sizes $p = 2, 4, 8$. The patch size must be a factor of 64 to evenly divide the total image domain. These patch sizes yield $1024$, $256$, and $64$ tokens respectively. Larger patch sizes embed more information per token, while smaller patches allow more tokens, and thus potentially richer relationships, to be extracted during self-attention.

In addition to patch size variation, three different model scales are used to determine internal training parameters such as embedding dimension, number of transformer blocks, and number of attention heads: DiT-Tiny, DiT-Small, and DiT-Base models are implemented as described in \cite{peebles_scalable_2023}. Larger models may have higher representational capacity but require larger datasets for sufficient training. The specifics of each model size are detailed in \cref{tbl:model_sizes}.

All models are trained on $T = 1000$ timesteps and subsampled at $T = 250$ timesteps for evaluation unless specified.

\section{Results}\label{sec: Results}
\begin{figure}
    \centering
    \includegraphics[width=1\linewidth, trim={32mm 32mm 25mm 30mm}, clip]{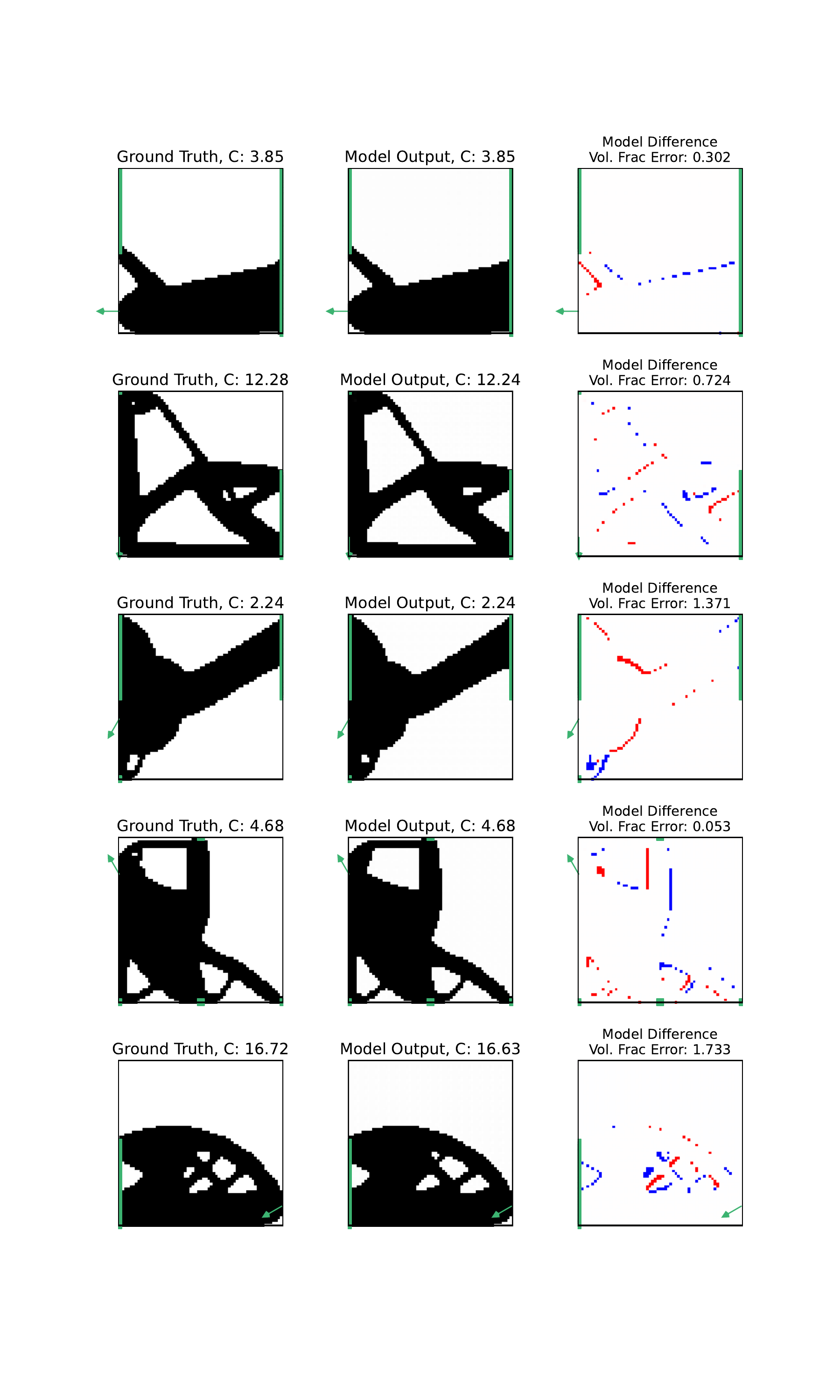}
    \caption{Results from the DiT-Small model with a patch size of 4, subsampled with 250 steps.}
    \label{fig:ditS_res}
\end{figure}

\begin{table*}[]
\resizebox{0.95\textwidth}{!}{%
\begin{tabular}{r|ccc|ccc|ccc}
& \multicolumn{9}{c}{\textbf{DiT Model}} \\
& \multicolumn{3}{c}{\textbf{Tiny}} & \multicolumn{3}{c}{\textbf{Small}} & \multicolumn{3}{c}{\textbf{Base}}
\\
        \cmidrule(lr){2-4} \cmidrule(lr){5-7} \cmidrule(lr){8-10}
\multicolumn{1}{r|}{\textbf{Patch Size}} &
  \textbf{2} &
  \textbf{4} &
  \textbf{8} &
  \textbf{2} &
  \textbf{4} &
  \textbf{8} &
  \textbf{2} &
  \textbf{4} &
  \textbf{8} \\
  \hline

Number   of Parameters&
  5.5M &
  5.5M &
  5.6M &
  32.6M &
  32.6M &
  32.7M &
  130M &
  130M &
  130M \\
Training Time (d:h:m) &
  1:05:31 &
  7:26 &
  5:16 &
  2:23:39 &
  17:48 &
  8:27 &
  3:00:00 &
  1:17:00 &
  12:53 \\
Sample Time (m:s) &
  3:35 &
  0:37 &
  0:09 &
  13:25 &
  1:31 &
  0:35 &
  42:35 &
  8:37 &
  2:03
\end{tabular}%
}
\caption{Model details for the DiT model suite. Training time is capped at 3 days or 1400 epochs. Sampling is performed with 250 step subsampling on a batch of 500 samples.}
\label{tbl:model_training}
\end{table*}

\begin{table*}
\resizebox{0.95\textwidth}{!}{%
\begin{tabular}{r|ccc|ccc|ccc}
& \multicolumn{9}{c}{\textbf{DiT Model}} \\
& \multicolumn{3}{c}{\textbf{Tiny}} & \multicolumn{3}{c}{\textbf{Small}} & \multicolumn{3}{c}{\textbf{Base}}
\\
        \cmidrule(lr){2-4} \cmidrule(lr){5-7} \cmidrule(lr){8-10}
\multicolumn{1}{r|}{\textbf{Patch Size}} &
  \textbf{2} &
  \textbf{4} &
  \textbf{8} &
  \textbf{2} &
  \textbf{4} &
  \textbf{8} &
  \textbf{2} &
  \textbf{4} &
  \textbf{8} \\
  \hline

Compliance Error (\%) &
  0.73 &
  1.21 &
  2.66 &
  0.35 &
  0.35 &
  0.50 &
  0.44 &
  0.40 &
  0.45 \\
Compliance Error Above 30\% (\%) &
  0.40 &
  0.40 &
  0.80 &
  0.20 &
  0.40 &
  0.40 &
  0.00 &
  0.00 &
  0.00 \\
Median Compliance Error (\%) &
  0.22 &
  0.32 &
  0.74 &
  0.16 &
  0.16 &
  0.24 &
  0.23 &
  0.20 &
  0.17 \\ \hline
Volume Fraction Error (\%) &
  0.83 &
  0.81 &
  0.92 &
  0.71 &
  0.64 &
  0.71 &
  1.02 &
  1.16 &
  0.79 \\
Load Discrepancy (\%) &
  0.00 &
  0.00 &
  0.20 &
  0.00 &
  0.00 &
  0.00 &
  0.00 &
  0.00 &
  0.00 \\
Floating   Material (\%) &
  0.40 &
  0.80 &
  5.40 &
  0.80 &
  0.20 &
  1.20 &
  0.80 &
  0.00 &
  1.00
\end{tabular}%
}
\caption{Full DiT model suite for various model sizes and patch sizes. Compliance error is presented as relative to the SIMP optimized dataset. Compliance error above 30\% represents the frequency of samples that have major structural differences from the ground truth, as measured by high compliance error.}
\label{tbl:model_results}
\end{table*}

The performance of the diffusion transformer framework was evaluated across multiple model scales and patch sizes to assess accuracy, computational efficiency, and sampling performance. \Cref{tbl:model_training} summarizes the number of trainable parameters, total training time, and average sampling time for each model configuration. \Cref{tbl:model_results} reports quantitative accuracy metrics including compliance error, compliance error above 30\%, median compliance error, volume fraction error, load discrepancy rate, and percentage of floating (disconnected) material.

Across all configurations, model size and patch size strongly influenced both training cost and inference speed. As expected, larger models such as DiT‑B contained up to 130M parameters and required longer training times but achieved marginally improved accuracy compared to smaller variants. The smallest configuration (DiT‑T) trained in significantly less time while maintaining sub 1\% compliance error. Patch size has a significant impact on training time as the number of tokens increases dramatically with smaller patch sizes. A patch size of 8 produces 64 tokens, a patch size of 4 produces 256 tokens, and a patch size of 2 produces 1024 tokens.

Sampling time scaled with the number of transformer layers and inversely with patch size. DiT‑S‑4 achieved a balance between accuracy and efficiency, completing sampling in under three minutes at 250 steps while maintaining high fidelity in predicted topologies. Increasing patch size reduced token count and accelerated inference but slightly degraded compliance error.

All models achieved low compliance errors relative to ground-truth SIMP results. The best-performing configurations reached mean compliance errors below 0.4\%, with median errors around 0.16\%. Even the smallest DiT variants maintained compliance errors under 3\%, outperforming previously reported CNN- or U-Net-based diffusion approaches that typically exhibit 4\% deviation from optimal topologies.

Volume fraction predictions remained within 1\% of target values across all models, demonstrating that the AdaLN conditioning effectively captured global design constraints such as prescribed material usage. Load discrepancy was negligible ($<$0.2\%), confirming that load location was respected during generation.

The percentage of floating or disconnected material regions was also low ($<$1\% for most models), indicating that generated structures retained physical connectivity consistent with valid mechanical designs.

\subsection{Subsampling}\label{subsec: Subsample}
\begin{table}[]
\resizebox{0.5\textwidth}{!}{%
\begin{tabular}{c|r@{:}l}
\hline
\multicolumn{1}{c}{\textbf{Number of Steps}} &
  \multicolumn{2}{c}{\textbf{Sample time (m:s)}} \\ \hline
        1000 &\hspace{0.5cm} 6&10 \\
        250  &1&31  \\
        100  &0&36  \\
        25   &0&09.1 \\
        10   &0&03.6 \\
        5    &0&02.22                    
\end{tabular}%
}
\caption{Sampling time for the DiT-S-4 model for different sampling steps. 500 samples are generated in one batch. Sampling time does not include model or batch initiation.}
\label{tbl:model_subsample_speed}
\end{table}

\begin{table*}[]
\centering
\resizebox{0.85\textwidth}{!}{%
\begin{tabular}{r|llllll}

\textbf{Number of Steps} &
  \textbf{1000} &
  \textbf{250} &
  \textbf{100} &
  \textbf{25} &
  \textbf{10} &
  \textbf{5} \\ \hline
Compliance   Error (\%) &
  0.40 &
  0.33 &
  0.38 &
  0.40 &
  0.40 &
  0.35 \\
Compliance Error Above 30\% (\%) &
  0.40 &
  0.20 &
  0.20 &
  0.40 &
  0.20 &
  0.20 \\
Median Compliance Error (\%) &
  0.15 &
  0.15 &
  0.16 &
  0.16 &
  0.13 &
  0.16 \\
  \hline
Volume Fraction Error (\%)&
  0.65 &
  0.63 &
  0.62 &
  0.63 &
  0.57 &
  0.56 \\
Load Discrepancy (\%)&
  0.00 &
  0.00 &
  0.00 &
  0.00 &
  0.00 &
  0.00 \\
Floating Material (\%)&
  0.00 &
  0.40 &
  0.20 &
  0.40 &
  1.40 &
  3.20 \\
\end{tabular}%
}
\caption{Results from subsampling tests, showing how the DiT-S-4 model handles sampling at 1000, 250, 100, 25, 10 and 5 steps}
\label{tbl:model_subsample_results}
\end{table*}

To evaluate the effect of DDIM undersampling on generation speed and accuracy, the DiT-S-4 model was tested with progressively fewer diffusion steps during inference (1000, 250, 100, 25, 10, and 5 steps).

\Cref{tbl:model_subsample_speed} shows the sampling time for the DiT-S-4, subsampled at different timestep counts. \Cref{tbl:model_subsample_results} shows the resulting accuracies for each subsample batch, averaged over 500 samples. From these results we observe that reducing the number of sampling steps produced substantial improvements in runtime while maintaining near-identical structural accuracy down to ten steps. As far as five steps (the fastest configuration) the model preserved overall topology with only a slight increase in compliance error. Sampling time decreased from over six minutes at full resolution (1000 steps) to less than three seconds at five steps. 

These results demonstrate that deterministic DDIM sampling enables real-time or near-real-time topology generation without significant degradation in quality - an essential feature for integration into interactive CAD workflows.

The diffusion transformer framework achieved sub-percent compliance errors across most configurations while drastically reducing inference time compared with traditional optimization methods that require iterative FEA solutions for each design case—typically several minutes per sample. The hybrid conditioning approach effectively integrated both global problem descriptors through AdaLN modulation and local mechanical context through stress/strain field concatenation, enabling accurate topology prediction even under diverse boundary conditions and load configurations.

Undersampled DDIM inference further demonstrated that high-quality topologies could be generated within seconds using as few as five denoising steps—representing a major step toward real-time structural optimization capabilities suitable for interactive design environments.

\subsection{Stress and Strain Validation}\label{subsec: SSValidation}
\begin{figure}
    \centering
    \includegraphics[width=1\linewidth, trim={0mm 0mm 0mm 0mm}, clip]{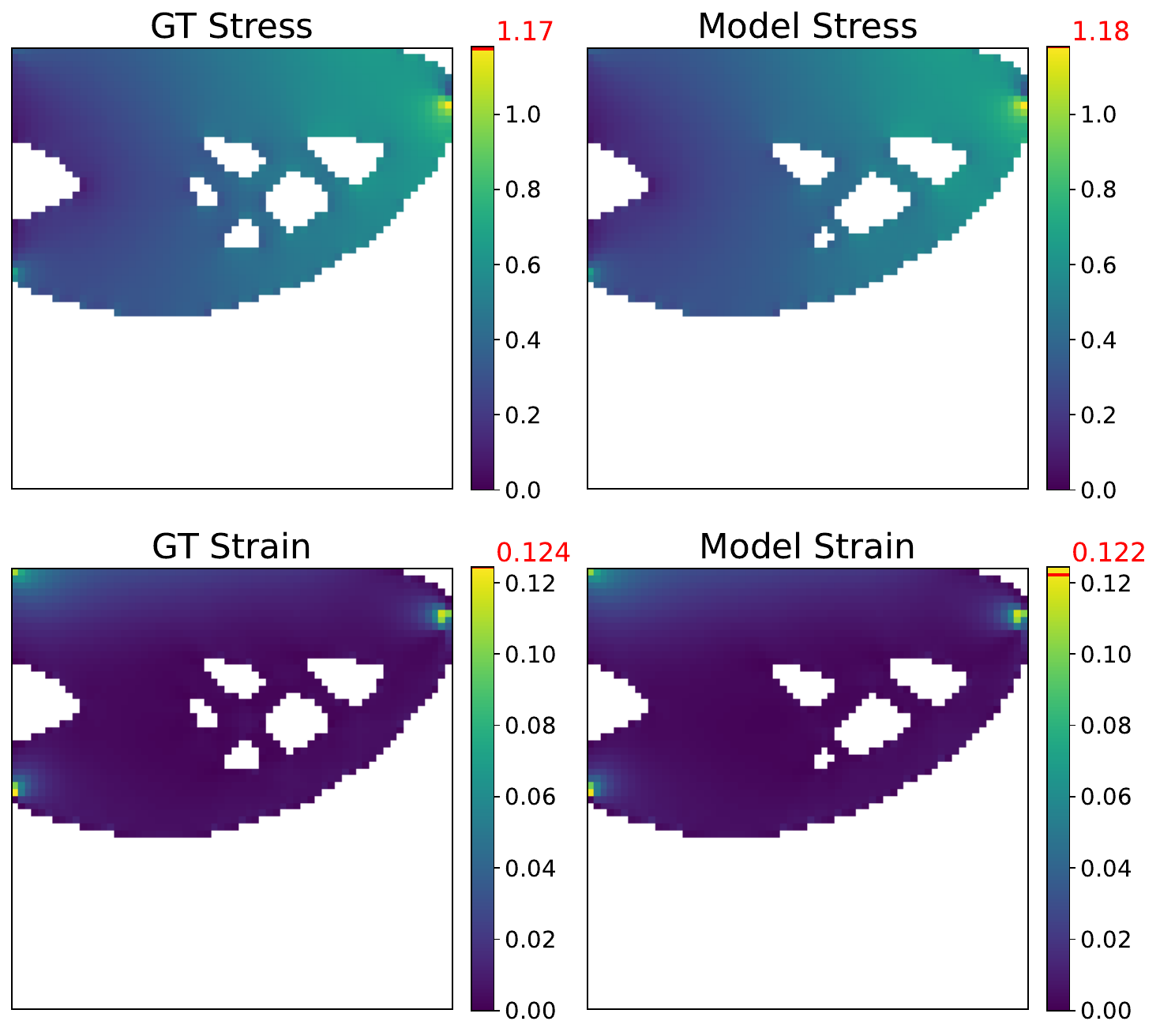}
    \caption{Stress and strain over a DiT-S-4 sample and ground truth. The maximums for stress and strain are marked on the color bar. Sample is from the DiT-Small-8 model.}
    \label{fig:stress_strain_example}
\end{figure}

\begin{figure}[h]
    \centering
    \includegraphics[width=0.8\linewidth, trim={10mm 6mm 20mm 0mm}]{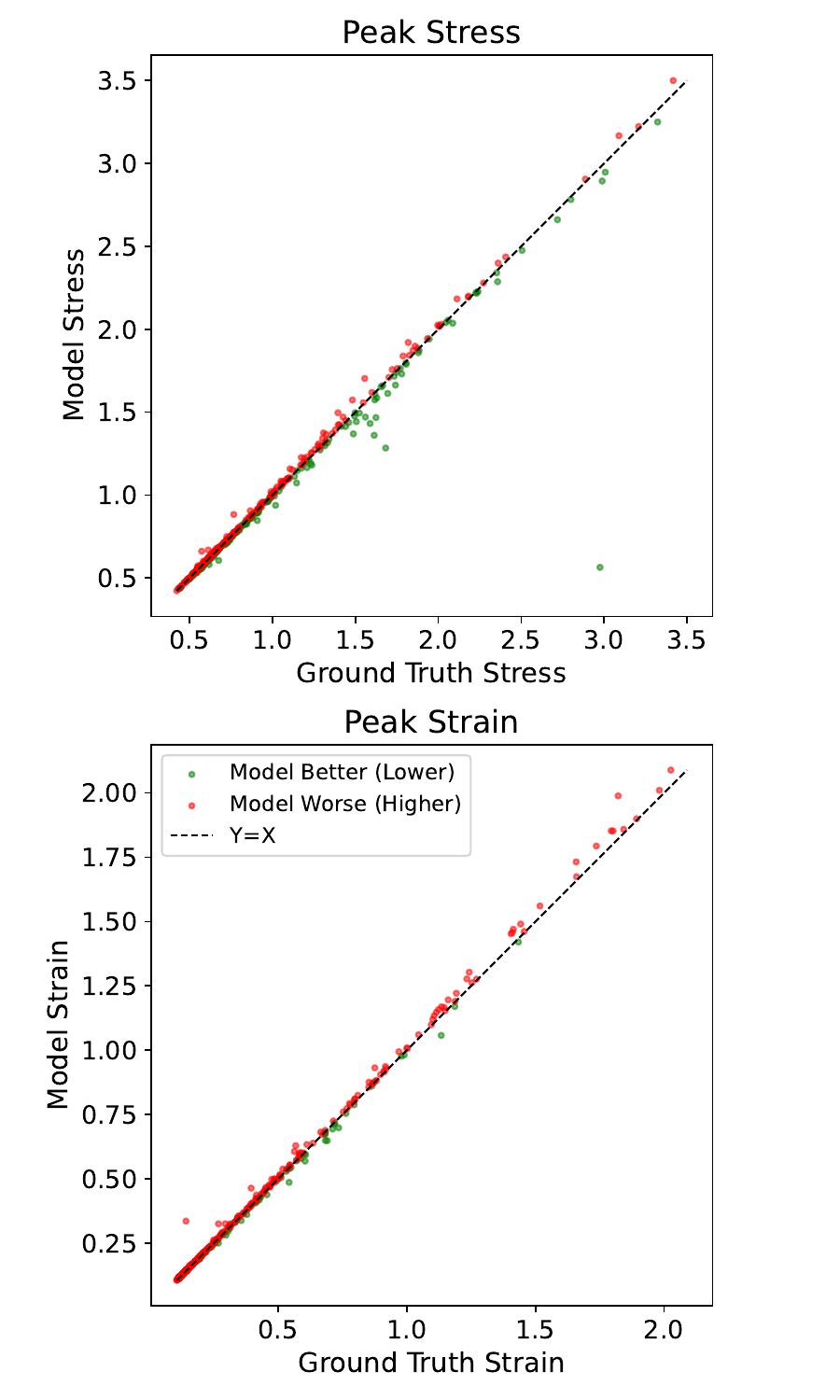}
    \caption{Scatter plots comparing peak von Mises stress (top) and peak strain energy density (bottom) between DiT‑generated topologies and ground‑truth SIMP results for 500 test samples. The dashed line indicates perfect agreement. Samples are generated from the DiT-S-4 model with 250 timesteps.}
    \label{fig:stress_strain_scatter}
\end{figure}

\begin{figure}[h]
    \centering
    \includegraphics[width=\linewidth]{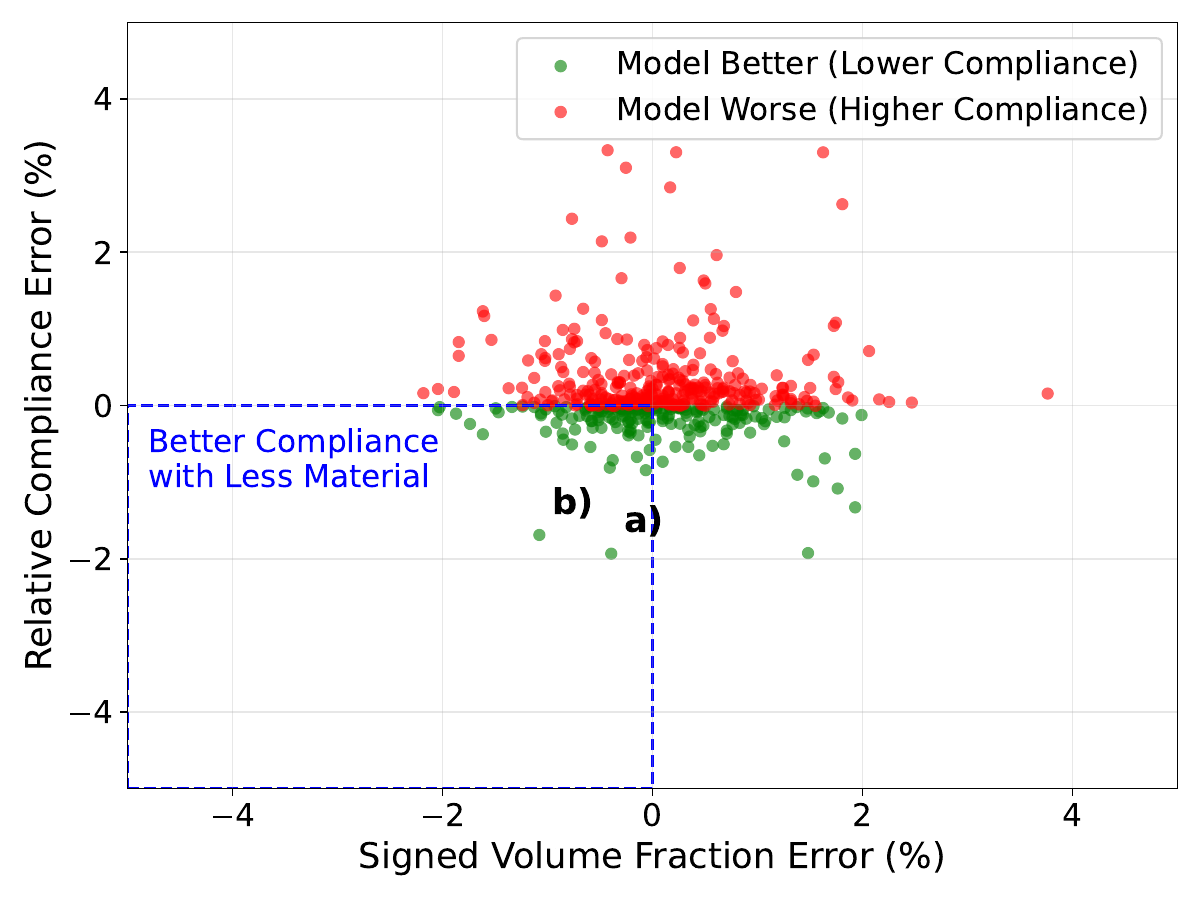}
    \caption{Scatter plot showing relationship between true compliance error and true volume fraction error for 500 validation samples. Samples are generated from the DiT-S-4 model with 250 timesteps. Two samples marked a) and b) and are shown in \cref{fig:comp_vol_examples}}
    \label{fig:comp_vol_corr}
\end{figure}

\begin{figure}[bh]
    \centering
    \begin{subfigure}[b]{\linewidth}
        \caption{\label{fig:comp_vol_examples_a}}
        \includegraphics[width=1\linewidth, trim={40mm, 105mm, 26mm, 12mm}, clip]{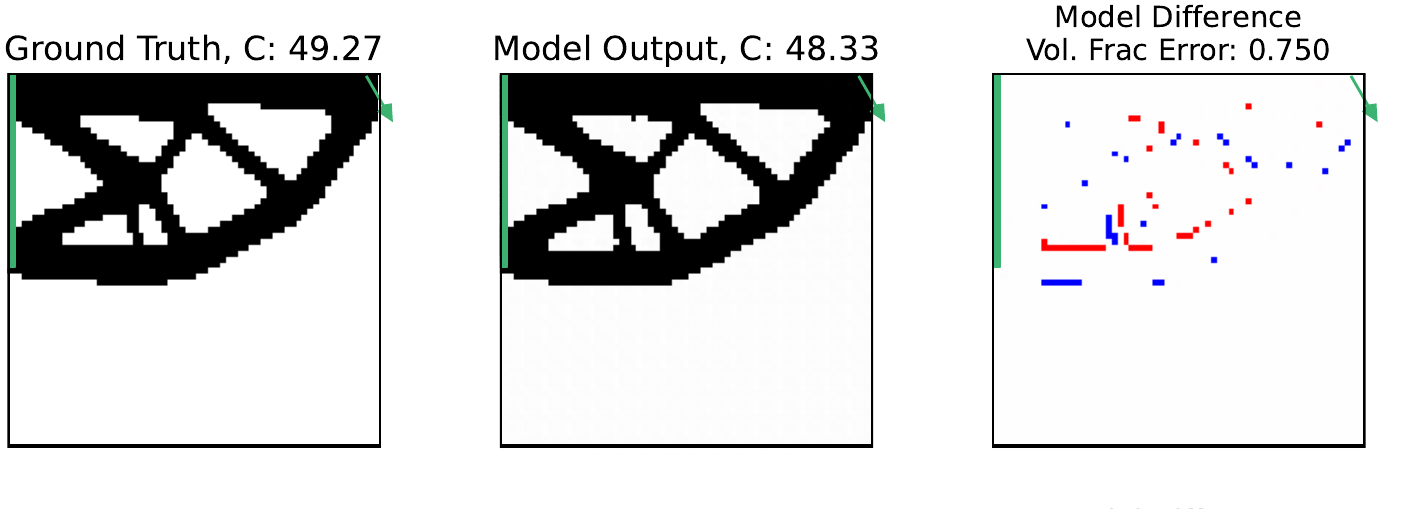}
    \end{subfigure}
    
    \begin{subfigure}[b]{\linewidth}
        \caption{\label{fig:comp_vol_examples_b}}
        \includegraphics[width=1\linewidth, trim={40mm, 15mm, 26mm, 95mm}, clip]{DiT-S_4_better_samples_plot.pdf}
    \end{subfigure}
    
    \caption{Samples from DiT-S-4 which perform better than ground truth in both compliance and volume fraction.}
    \label{fig:comp_vol_examples}
\end{figure}

While compliance provides a measure of global structural performance, localized stress and strain distributions are indicators of the viability of topologies. To assess the physical fidelity of samples, we performed a validation study comparing the predicted stress and strain in structures from the DiT-S-4 model against ground truth samples.

For each of 500 validation samples, FEA was performed on both samples. The maximum von Mises stress $\sigma_max$ and maximum strain energy density $\epsilon_max$ were extracted from the analysis. These peak values capture the most critical elements in the analysis and give insight into the reliability of optimized structures. \cref{fig:stress_strain_scatter} shows scatter plots of model versus ground truth peak stresses and strains. Each point represents one sample, with the dashed diagonal line indicating perfect correspondence. Points below the line are model topologies which perform better than ground truth. Points above perform worse. An example stress and strain distribution is shown in \cref{fig:stress_strain_example}.

Both stress and strain show tight packing around the diagonal, indicating that the DiT produces samples which have similar critical elements as the ground truth. The DiT model not only reproduces global compliance accurately but also maintains accurate local mechanical response.

\subsection{Compliance-Volume Correlation}\label{subsec: CVCorrelation}
An additional analysis was performed to investigate whether compliance error correlates with deviations in volume fraction. Applying more material onto a topology can improve the compliance error at the cost of volume error. To determine if the DiT model is adding extra material to gain compliance performance, we assess the correlation between compliance error and volume fraction error. Unlike previous sections where absolute errors were reported, here we use signed errors to capture both over- and under-performance relative to ground truth.

As shown in \cref{fig:comp_vol_corr}, most samples lie within the positive compliance error region, indicating slightly higher compliance than ground truth; however, a substantial number exhibit negative compliance errors. No strong correlation is observed between volume fraction and compliance error, suggesting that improved performance is not achieved simply by adding excess material.

Notably, several samples appear in the negative-negative quadrant of the scatter plot, indicating both lower compliance and lower material usage than their corresponding SIMP designs. Two representative cases from this region are highlighted for closer inspection in \cref{fig:comp_vol_examples}.

\section{Conclusion}\label{sec:Conclusion}
This work introduced a DiT framework for data‑driven structural topology optimization, combining the representational capacity of diffusion models with the long-range relationship capture of transformer models. By integrating spatial conditioning through stress and strain fields and global conditioning via adaLN, our model effectively captures both local mechanical behavior and global design constraints.

The DiT framework achieves compliance errors well below 1\% relative to SIMP-optimized ground truth while maintaining accurate volume fractions and structural connectivity. Deterministic DDIM sampling further reduced inference time by several orders of magnitude, allowing high‑fidelity topology generation in seconds using as few as five denoising steps, demonstrating clear potential for real-time or interactive design applications. 

Our hybrid-conditioned DiT framework represents a scalable, physics-aware alternative to traditional iterative optimization methods. Its ability to generate near-optimal structures rapidly makes it well suited for integration in workflows where designers can explore multiple configurations interactively. Future work can easily extend this approach to three-dimensional domains, multi-load, and multi-material problems.

\section*{Declarations}
\noindent \textbf{Conflict of interest} On behalf of all authors, the corresponding author states that there is no conflict of interest.

\noindent \textbf{Funding} This work has been financially supported by the Institute of Digital Engineering - USA.

\noindent \textbf{Author contributions} Conceptualization: Srijan Das, Alireza Tabarraei; Methodology: Aaron Lutheran; Formal analysis and investigation: Aaron Lutheran; Writing - original draft preparation: Aaron Lutheran; Writing - review and editing: Srijan Das, Alireza Tabarraei; Supervision, Srijan Das, Alireza Tabarraei.

\noindent \textbf{Ethics approval and Consent to participate} Not applicable for this work.

\noindent \textbf{Data Availability} Dataset will be made available on request to the corresponding author.

\noindent \textbf{Replication of results} Replication material, including model parameters, and code, are available on request to the corresponding author.

\nocite{*}

\bibliography{DiffusionTransformer}

@article{nie_topologygan_2021,
	title = {{TopologyGAN}: {Topology} {Optimization} {Using} {Generative} {Adversarial} {Networks} {Based} on {Physical} {Fields} {Over} the {Initial} {Domain}},
	volume = {143},
	issn = {1050-0472},
	shorttitle = {{TopologyGAN}},
	number = {031715},
	journal = {Journal of Mechanical Design},
	author = {Nie, Zhenguo and Lin, Tong and Jiang, Haoliang and Kara, Levent Burak},
	month = feb,
	year = {2021},
}

@misc{maze_diffusion_2022,
	title = {Diffusion {Models} {Beat} {GANs} on {Topology} {Optimization}},
	publisher = {arXiv},
	author = {Mazé, François and Ahmed, Faez},
	month = dec,
	year = {2022},
}

@misc{salimans_progressive_2022,
	title = {Progressive {Distillation} for {Fast} {Sampling} of {Diffusion} {Models}},
	publisher = {arXiv},
	author = {Salimans, Tim and Ho, Jonathan},
	month = jun,
	year = {2022},
}

@misc{giannone_diffusing_2023,
	title = {Diffusing the {Optimal} {Topology}: {A} {Generative} {Optimization} {Approach}},
	shorttitle = {Diffusing the {Optimal} {Topology}},
	publisher = {arXiv},
	author = {Giannone, Giorgio and Ahmed, Faez},
	month = mar,
	year = {2023},
}

@misc{song_denoising_2022,
	title = {Denoising {Diffusion} {Implicit} {Models}},
	publisher = {arXiv},
	author = {Song, Jiaming and Meng, Chenlin and Ermon, Stefano},
	month = oct,
	year = {2022},
}

@misc{nichol_improved_2021,
	title = {Improved {Denoising} {Diffusion} {Probabilistic} {Models}},
	publisher = {arXiv},
	author = {Nichol, Alex and Dhariwal, Prafulla},
	month = feb,
	year = {2021},
}

@misc{hong_improving_2023,
	title = {Improving {Sample} {Quality} of {Diffusion} {Models} {Using} {Self}-{Attention} {Guidance}},
	publisher = {arXiv},
	author = {Hong, Susung and Lee, Gyuseong and Jang, Wooseok and Kim, Seungryong},
	month = aug,
	year = {2023},
}

@misc{yang_diffusion_2023,
	title = {Diffusion {Models}: {A} {Comprehensive} {Survey} of {Methods} and {Applications}},
	shorttitle = {Diffusion {Models}},
	publisher = {arXiv},
	author = {Yang, Ling and Zhang, Zhilong and Song, Yang and Hong, Shenda and Xu, Runsheng and Zhao, Yue and Zhang, Wentao and Cui, Bin and Yang, Ming-Hsuan},
	month = oct,
	year = {2023},
}

@misc{song_solving_2023,
	title = {Solving {Inverse} {Problems} with {Latent} {Diffusion} {Models} via {Hard} {Data} {Consistency}},
	publisher = {arXiv},
	author = {Song, Bowen and Kwon, Soo Min and Zhang, Zecheng and Hu, Xinyu and Qu, Qing and Shen, Liyue},
	month = oct,
	year = {2023},
}

@article{andreassen_efficient_2011,
	title = {Efficient topology optimization in {MATLAB} using 88 lines of code},
	volume = {43},
	issn = {1615-1488},
	language = {en},
	number = {1},
	journal = {Structural and Multidisciplinary Optimization},
	author = {Andreassen, Erik and Clausen, Anders and Schevenels, Mattias and Lazarov, Boyan S. and Sigmund, Ole},
	month = jan,
	year = {2011},
	pages = {1--16},
}

@article{white_computational_2019,
	title = {A computational study of symmetry and well-posedness of structural topology optimization},
	volume = {59},
	issn = {1615-1488},
	language = {en},
	number = {3},
	journal = {Structural and Multidisciplinary Optimization},
	author = {White, Daniel A. and Voronin, Alexey},
	month = mar,
	year = {2019},
	pages = {759--766},
}

@article{sigmund_topology_2013,
	title = {Topology optimization approaches},
	volume = {48},
	issn = {1615-1488},
	language = {en},
	number = {6},
	journal = {Structural and Multidisciplinary Optimization},
	author = {Sigmund, Ole and Maute, Kurt},
	month = dec,
	year = {2013},
	pages = {1031--1055},
}

@article{bendsoe_optimal_1989,
	title = {Optimal shape design as a material distribution problem},
	volume = {1},
	issn = {1615-1488},
	language = {en},
	number = {4},
	journal = {Structural optimization},
	author = {Bendsøe, M. P.},
	month = dec,
	year = {1989},
	pages = {193--202},
}

@article{allaire_level-set_2002,
	title = {A level-set method for shape optimization},
	volume = {334},
	issn = {1631-073X},
	number = {12},
	journal = {Comptes Rendus Mathematique},
	author = {Allaire, Grégoire and Jouve, François and Toader, Anca-Maria},
	month = jan,
	year = {2002},
	pages = {1125--1130},
}

@misc{rombach_high-resolution_2022,
	title = {High-{Resolution} {Image} {Synthesis} with {Latent} {Diffusion} {Models}},
	publisher = {arXiv},
	author = {Rombach, Robin and Blattmann, Andreas and Lorenz, Dominik and Esser, Patrick and Ommer, Björn},
	month = apr,
	year = {2022},
}

@article{shishir_multimaterials_2024,
	title = {Multi–materials topology optimization using deep neural network for coupled thermo–mechanical problems},
	volume = {291},
	issn = {0045-7949},
	journal = {Computers \& Structures},
	author = {Shishir, Md. Imrul Reza and Tabarraei, Alireza},
	month = jan,
	year = {2024},
	pages = {107218},
}

@article{zhang_research_2025,
	title = {Research on multi-stage topology optimization method based on latent diffusion model},
	volume = {63},
	issn = {1474-0346},
	journal = {Advanced Engineering Informatics},
	author = {Zhang, Wei and Zhao, Guodong and Su, Lijie},
	month = jan,
	year = {2025},
	pages = {102966},
}

@article{yin_dynamically_2024,
	title = {Dynamically configured physics-informed neural network in topology optimization applications},
	volume = {426},
	issn = {0045-7825},
	journal = {Computer Methods in Applied Mechanics and Engineering},
	author = {Yin, Jichao and Wen, Ziming and Li, Shuhao and Zhang, Yaya and Wang, Hu},
	month = jun,
	year = {2024},
	pages = {117004},
}

@article{chandrasekhar_tounn_2021,
	title = {{TOuNN}: {Topology} {Optimization} using {Neural} {Networks}},
	volume = {63},
	issn = {1615-1488},
	shorttitle = {{TOuNN}},
	language = {en},
	number = {3},
	journal = {Structural and Multidisciplinary Optimization},
	author = {Chandrasekhar, Aaditya and Suresh, Krishnan},
	month = mar,
	year = {2021},
	pages = {1135--1149},
}

@misc{banga_3d_2018,
	title = {{3D} {Topology} {Optimization} using {Convolutional} {Neural} {Networks}},
	publisher = {arXiv},
	author = {Banga, Saurabh and Gehani, Harsh and Bhilare, Sanket and Patel, Sagar and Kara, Levent},
	month = aug,
	year = {2018},
	note = {arXiv:1808.07440 [cs]},
}

@article{jeong_physics-informed_2023,
	title = {A {Physics}-{Informed} {Neural} {Network}-based {Topology} {Optimization} ({PINNTO}) framework for structural optimization},
	volume = {278},
	issn = {0141-0296},
	journal = {Engineering Structures},
	author = {Jeong, Hyogu and Bai, Jinshuai and Batuwatta-Gamage, C. P. and Rathnayaka, Charith and Zhou, Ying and Gu, YuanTong},
	month = mar,
	year = {2023},
	pages = {115484},
}

@article{behzadi_real-time_2021,
	title = {Real-{Time} {Topology} {Optimization} in {3D} via {Deep} {Transfer} {Learning}},
	volume = {135},
	issn = {0010-4485},
	journal = {Computer-Aided Design},
	author = {Behzadi, Mohammad Mahdi and Ilieş, Horea T.},
	month = jun,
	year = {2021},
	pages = {103014},
}

@article{mukherjee_accelerating_2021,
	title = {Accelerating {Large}-scale {Topology} {Optimization}: {State}-of-the-{Art} and {Challenges}},
	volume = {28},
	issn = {1886-1784},
	shorttitle = {Accelerating {Large}-scale {Topology} {Optimization}},
	language = {en},
	number = {7},
	journal = {Archives of Computational Methods in Engineering},
	author = {Mukherjee, Sougata and Lu, Dongcheng and Raghavan, Balaji and Breitkopf, Piotr and Dutta, Subhrajit and Xiao, Manyu and Zhang, Weihong},
	month = dec,
	year = {2021},
	pages = {4549--4571},
}

@misc{peebles_scalable_2023,
	title = {Scalable {Diffusion} {Models} with {Transformers}},
	publisher = {arXiv},
	author = {Peebles, William and Xie, Saining},
	month = mar,
	year = {2023},
}

@article{vaswani_attention_2017,
  title={Attention is all you need},
  author={Vaswani, Ashish and Shazeer, Noam and Parmar, Niki and Uszkoreit, Jakob and Jones, Llion and Gomez, Aidan N and Kaiser, {\L}ukasz and Polosukhin, Illia},
  journal={Advances in neural information processing systems},
  volume={30},
  year={2017}
}

@misc{perez_film_2017,
	title = {{FiLM}: {Visual} {Reasoning} with a {General} {Conditioning} {Layer}},
	shorttitle = {{FiLM}},
	publisher = {arXiv},
	author = {Perez, Ethan and Strub, Florian and Vries, Harm de and Dumoulin, Vincent and Courville, Aaron},
	month = dec,
	year = {2017},
}

@article{guo_adaln_2022,
	title = {{AdaLN}: {A} {Vision} {Transformer} for {Multidomain} {Learning} and {Predisaster} {Building} {Information} {Extraction} from {Images}},
	volume = {36},
	shorttitle = {{AdaLN}},
	language = {en},
	number = {5},
	journal = {Journal of Computing in Civil Engineering},
	author = {Guo, Yunhui and Wang, Chaofeng and Yu, Stella X. and McKenna, Frank and Law, Kincho H.},
	month = sep,
	year = {2022},
	pages = {04022024},
}

@article{shin_topology_2023,
	title = {Topology optimization via machine learning and deep learning: a review},
	volume = {10},
	issn = {2288-5048},
	shorttitle = {Topology optimization via machine learning and deep learning},
	number = {4},
	journal = {Journal of Computational Design and Engineering},
	author = {Shin, Seungyeon and Shin, Dongju and Kang, Namwoo},
	month = aug,
	year = {2023},
	pages = {1736--1766},
}

@article{rong_structural_2022,
	title = {Structural topology optimization with an adaptive design domain},
	volume = {389},
	issn = {0045-7825},
	journal = {Computer Methods in Applied Mechanics and Engineering},
	author = {Rong, Yi and Zhao, Zi-Long and Feng, Xi-Qiao and Xie, Yi Min},
	month = feb,
	year = {2022},
	pages = {114382},
}

@incollection{bendsoe_topology_2004,
	address = {Berlin, Heidelberg},
	title = {Topology optimization by distribution of isotropic material},
	isbn = {978-3-662-05086-6},
	language = {en},
	booktitle = {Topology {Optimization}: {Theory}, {Methods}, and {Applications}},
	publisher = {Springer},
	author = {Bendsøe, Martin P. and Sigmund, Ole},
	editor = {Bendsøe, Martin P. and Sigmund, Ole},
	year = {2004},
	pages = {1--69},
}

@article{wang_projection_2011,
	title = {On projection methods, convergence and robust formulations in topology optimization},
	volume = {43},
	issn = {1615-1488},
	language = {en},
	number = {6},
	journal = {Structural and Multidisciplinary Optimization},
	author = {Wang, Fengwen and Lazarov, Boyan Stefanov and Sigmund, Ole},
	month = jun,
	year = {2011},
	pages = {767--784},
}

@article{zhu_filter_2015,
	title = {Filter the shape sensitivity in level set-based topology optimization methods},
	volume = {51},
	issn = {1615-1488},
	language = {en},
	number = {5},
	journal = {Structural and Multidisciplinary Optimization},
	author = {Zhu, Benliang and Zhang, Xianmin and Fatikow, Sergej},
	month = may,
	year = {2015},
	pages = {1035--1049},
}

@article{lazarov_length_2016,
	title = {Length scale and manufacturability in density-based topology optimization},
	volume = {86},
	issn = {1432-0681},
	language = {en},
	number = {1},
	journal = {Archive of Applied Mechanics},
	author = {Lazarov, Boyan S. and Wang, Fengwen and Sigmund, Ole},
	month = jan,
	year = {2016},
	pages = {189--218},
}

@article{luo_level_2008,
	title = {A level set-based parameterization method for structural shape and topology optimization},
	volume = {76},
	issn = {1097-0207},
	language = {en},
	number = {1},
	journal = {International Journal for Numerical Methods in Engineering},
	author = {Luo, Zhen and Wang, Michael Yu and Wang, Shengyin and Wei, Peng},
	year = {2008},
	pages = {1--26},
}

@article{tang_topology_2024,
	title = {Topology {Optimization}: {A} {Review} for {Structural} {Designs} {Under} {Statics} {Problems}},
	volume = {17},
	copyright = {http://creativecommons.org/licenses/by/3.0/},
	issn = {1996-1944},
	shorttitle = {Topology {Optimization}},
	language = {en},
	number = {23},
	journal = {Materials},
	author = {Tang, Tianshu and Wang, Leijia and Zhu, Mingqiao and Zhang, Huzhi and Dong, Jiarui and Yue, Wenhui and Xia, Hui},
	month = jan,
	year = {2024},
	pages = {5970},
}

@article{zheng_generating_2021,
	title = {Generating three-dimensional structural topologies via a {U}-{Net} convolutional neural network},
	volume = {159},
	issn = {0263-8231},
	journal = {Thin-Walled Structures},
	author = {Zheng, Shuai and He, Zhenzhen and Liu, Honglei},
	month = feb,
	year = {2021},
	pages = {107263},
}

@article{wang_structural_2021,
	title = {Structural topology optimization considering both performance and manufacturability: strength, stiffness, and connectivity},
	volume = {63},
	issn = {1615-1488},
	shorttitle = {Structural topology optimization considering both performance and manufacturability},
	language = {en},
	number = {3},
	journal = {Structural and Multidisciplinary Optimization},
	author = {Wang, Chao and Xu, Bin and Duan, Zunyi and Rong, Jianhua},
	month = mar,
	year = {2021},
	pages = {1427--1453},
}

@article{cuilliere_towards_2014,
	title = {Towards the {Integration} of {Topology} {Optimization} into the {CAD} {Process}},
	volume = {11},
	issn = {null},
	number = {2},
	journal = {Computer-Aided Design and Applications},
	author = {Cuillière, Jean-Christophe and Francois, Vincent and Drouet, Jean-Marc},
	month = mar,
	year = {2014},
	pages = {120--140},
}

@article{croitoru_diffusion_2023,
	title = {Diffusion {Models} in {Vision}: {A} {Survey}},
	volume = {45},
	issn = {1939-3539},
	shorttitle = {Diffusion {Models} in {Vision}},
	number = {9},
	journal = {IEEE Transactions on Pattern Analysis and Machine Intelligence},
	author = {Croitoru, Florinel-Alin and Hondru, Vlad and Ionescu, Radu Tudor and Shah, Mubarak},
	month = sep,
	year = {2023},
	pages = {10850--10869},
}

@article{dhariwal_diffusion_2021,
  title={Diffusion models beat gans on image synthesis},
  author={Dhariwal, Prafulla and Nichol, Alexander},
  journal={Advances in neural information processing systems},
  volume={34},
  pages={8780--8794},
  year={2021}
}

@misc{lutheran_latent_2025,
	title = {Latent {Space} {Diffusion} for {Topology} {Optimization}},
	publisher = {arXiv},
	author = {Lutheran, Aaron and Das, Srijan and Tabarraei, Alireza},
	month = aug,
	year = {2025},
}

@Inbook{hajela1993genetic,
    author="Hajela, P.
    and Lee, E.
    and Lin, C.-Y.",
    editor="Bends{\o}e, Martin Philip
    and Soares, Carlos A. Mota",
    title="Genetic Algorithms in Structural Topology Optimization",
    bookTitle="Topology Design of Structures",
    year="1993",
    publisher="Springer Netherlands",
    address="Dordrecht",
    pages="117--133",
    isbn="978-94-011-1804-0",
}

@article{luh2011binary,
  title={A binary particle swarm optimization for continuum structural topology optimization},
  author={Luh, Guan-Chun and Lin, Chun-Yi and Lin, Yu-Shu},
  journal={Applied Soft Computing},
  volume={11},
  number={2},
  pages={2833--2844},
  year={2011},
  publisher={Elsevier}
}

@article{garcia2011hybrid,
  title={A hybrid topology optimization methodology combining simulated annealing and SIMP},
  author={Garcia-Lopez, NP and Sanchez-Silva, M and Medaglia, AL and Chateauneuf, A},
  journal={Computers \& structures},
  volume={89},
  number={15-16},
  pages={1512--1522},
  year={2011},
  publisher={Elsevier}
}

@article{cui2002topology,
  title={Topology optimization for maximum natural frequency using simulated annealing and morphological representation},
  author={Cui, Guang Yu and Tai, Kang and Wang, Bo Ping},
  journal={AIAA journal},
  volume={40},
  number={3},
  pages={586--589},
  year={2002}
}

@article{lutheran2026physics,
  title={Physics-Informed Transformer for Real-Time High-Fidelity Topology Optimization},
  author={Lutheran, Aaron and Das, Srijan and Tabarraei, Alireza},
  journal={arXiv preprint arXiv:2604.03522},
  year={2026}
}

@article{SHISHIR2024107218,
    title = {Multi–materials topology optimization using deep neural network for coupled thermo–mechanical problems},
    journal = {Computers \& Structures},
    volume = {291},
    pages = {107218},
    year = {2024},
    issn = {0045-7949},
    author = {Md. Imrul Reza Shishir and Alireza Tabarraei},
}

\end{document}